\documentclass[amsmath,amssymb,twocolumn]{revtex4-2}
\usepackage{bm}
\usepackage{epsfig}
\usepackage{amsmath}
\usepackage{color}
\newcommand{\nix}[1]{}

\usepackage[unicode=true,colorlinks=true,urlcolor=blue,citecolor=blue]{hyperref}

\begin{document}

%mapping
%\title{Spin imaging of Poiseuille electronic flow in ultrahigh-mobility 2DEG}
\title{Spin imaging of Poiseuille flow of viscous electronic fluid}
\author{K.S.~Denisov}
%\affiliation{Ioffe Institute, 194021 St.Petersburg, Russia}	
\email{denisokonstantin@gmail.com}
\author{K.A~Baryshnikov}
%	\email{denisokonstantin@gmail.com}
%\affiliation{Ioffe Institute, 194021 St.Petersburg, Russia} 
\author{P.S.~Alekseev}
\affiliation{
Ioffe Institute, 194021 St.Petersburg, Russia} 
%\email{?}
\begin{abstract}
%\textcolor{red}{two-dimensional}
Recent progress in fabricating high-quality conductors with small densities of defects has initiated the studies  of 
% phenomena emerging due to 
 %the formation of 
the viscous electron fluid and has motivated the search for the evidences of the hydrodynamic regime of electron transport. 
%%that allows one to 
In this work we come up with the spin imaging technique allowing us to 
attest to the emergence of electron hydrodynamic flows.
%%%%%%%%hydrodynamics in viscous electron fluids when the electric current flows. % 
%%%%%attest to the formation of viscous electron fluid by detecting the spatial pattern of the injected spin polarization %across the sample 
%by means of detecting the spatial pattern %of electron fluid 
%of the spin polarization 
%across the sample 
%upon the application of the dragging electric field. 
Based on numerical calculations  
we demonstrate that the injected electron spin density is inhomogeneous across the channel when the viscous electron fluid forms the   Poiseuille flow.  
% obeys the distribution, % in the hydrodynamic regime of the electron transport,
 %with the particular shape controlled by the ratio of the spin diffusion length and the channel width. 
We also argue that 
the Hanle curves %%%probed for injected spin density 
 % measured
at different positions across the channel 
%in case of 
%%for an in-plane magnetic field 
%%%the application of 
acquire relative phase shifts resulting from the variation of the electron drift velocity 
in inhomogeneous hydrodynamic
%  inhomogeneous hydrodynamic 
flows.
The studied   effects  % paves the way towards
 can be employed to evidence and   
  study  the viscous electron fluid non-invasively.
%by local spin probes. 
%, 
% which can be used to independently evidence the formation of the viscous 
% electron fluid.
%%%as an independent evidence for the formation of the viscous electron fluid.
%which serves as an independent evidence for 
%in the Poiseuille regime, 
%which serves as an independent evidence for 
%the formation of inhomogeneous hydrodynamic flows of electron fluid. 

%relative
%An overwhelming increase in interest to hydrodynamic effects in electronic flows in ultra-high mobility two-dimensional materials points out the necessity of alternative tools to detect these kind of a regime. We suggest a new way of imaging of Poiseuille electronic flow by means of spin polarization detection across GaAs quantum wells. The method is based on the fact that Poiseuille-like distribution is formed for spin polarization when the hydrodynamical regime is stabled. 
%%%We prove this fact by calculations made with the true parameters of electron-electron collision times, spin relaxation times, spin diffusion coefficients and other parameters for ultra-high mobility GaAs quantum wells. 
%%%%%%%%%%We show that the application of in-plane magnetic fields yields the Hanle curves for spin flow in the center and at the boundary of quantum well to be shifted by a phase shift due to the difference of drift velocities in the hydrodynamical mode. This two facts make possible to approve and to map the Poiseuille flow of electrons in a cheap way from experimental point of view.
	
\end{abstract}

\date{\today}

\maketitle

\newpage

%\textit{Introduction.} 
In high-quality conductors with small densities of defects, electrons can form a viscous fluid at low temperatures due to frequent electron-electron collisions and/or elasticity effects from the inter-particle interaction. The charge  transport in such fluid
is carried out by inhomogeneous hydrodynamic flows, controlled by particular shapes of samples, 
while its
%%%The 
resistance 
becomes proportional to the viscosity coefficient.  
These ideas were first proposed and partially theoretically studied 
%%%%%%\textcolor{red}{many}
%%%%%%years ago 
for bulk metals with strong electron-phonon coupling~\cite{Gurzhi}. 
%\textcolor{blue}{, appeared ahead of its time.}
Recently, 
this topic has become of interest as 
%\textcolor{blue}{the interest to this issue has resumed, as}
the hydrodynamic regime of electron transport has been realized in high-quality samples of graphene~\cite{grahene, grahene_2, grahene_3, samaddar2021evidence, Levitov_et_al, profile_1, profile_2}, quasi-two dimensional metal PdCoO$_2$~\cite{Weyl_sem_1}, Weyl semimetal WP$_2$~\cite{Weyl_sem_2}, and high-mobility GaAs quantum wells~\cite{exps_neg_1, exps_neg_2, exps_neg_3, exps_neg_4, je_visc, Gusev_1, Gusev_2, Gusev_3, recent___1, recent___2,exp_ac_GaAs_1,exp_ac_GaAs_2,exp_ac_GaAs_3}. These experiments motivated many theoretical works (see, for example~\cite{ph_tr_num, Levitov_et_al_2, Lucas, eta_xy, we_3, Lucas_2,  recentest_, recentest___breathing_flow, recentest2, L_n_1, recentest3, vis_res, Khoo_Villadiego, future, Alekseev_Alekseeva,  future2, we_6}), 
which were aimed to formulation and 
search for 
the evidences of the hydrodynamic regime as well as to studying of various types and regimes of flows of the electron fluid.

\begin{figure}[htbp]
	\centering
	\includegraphics[width=0.4\textwidth]{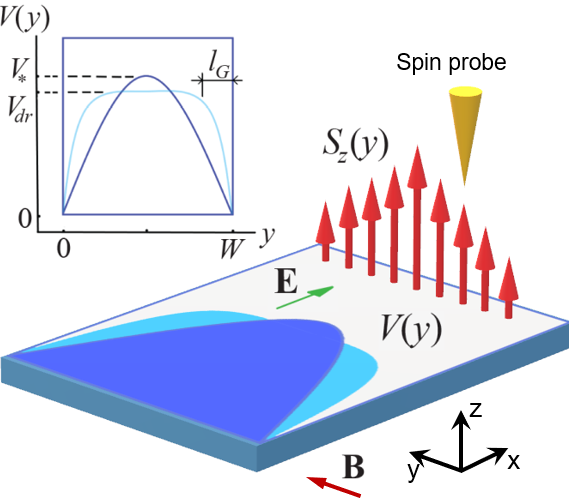}
	\caption{
    Scheme of the proposed spin imaging technique. 
	%%%%allowing to disclose an inhomogeneous distribution of the flow velocity $V(y)$ of an electron fluid upon dragging electric field ${\bm E}$. 
	Blue and bright blue curves depict the electron flow velocity $V(y)$ 
	under electric field ${\bm E}$ 
	for Poiseuille and 	Ohmic regimes, respectively.  % of the electric current, 
	%\textcolor{red}{ distribution}
	The spin polarization $S_z(y)$ (indicated by red arrows) 
	keeps an inhomogeneous structure at some distance from the injector 
	due to hydrodynamic flow of electrons 
	and can be detected by a local spin probe. 
	Applying an in-plane magnetic field ${\bm B}$ 
	one additionally detects the phase shifts in the Hanle curves tracked 
	at different positions across the sample 
	due to the variation of $V(y)$, here $V_\ast = V_x(W/2)$.
	} 
%%%% Kirill version
%%%%The principal scheme of the proposed spin imaging technique for	the inhomogeneous distribution of electronic velocities across the channel $V(y)$ due to dragging electric field ${\bm E}$ is depicted by blue and bright blue curves. The latter has some Gurzhi length $l_G$ smaller than the width of the channel $W$. 	As it is shown in this work, 	spin polarization $S_z(y)$ of electronic flow takes a Poiseuille-like form too,	which is shown by red arrows at some distance from the injection barrier 	and which can be resolved by local spin probe detection Applying an in-plane magnetic field ${\bm B}$ 	one measures the phase shifts of the Hanle curves inhomogeneous electron fluid flow.
\label{fig1}
\end{figure}

The evidences of formation of a viscous electron fluid are based, first, on an inhomogeneity  of space distributions of 
its flows, leading to the specific properties of observed sample resistances. The simplest of these properties is the cubic dependence of the conductance %\textcolor{red}{(reciprocal resistance)}
on the sample width. This dependence  was observed for the first time in~\cite{Weyl_sem_1} for stripes of PdCoO$_2$.  In samples of peculiar geometry of edges and contacts, whirlpools can appear,
%\textcolor{red}{similar to ones in water flows in rivers. }
similarly to water flows in rivers.
Herewith opposite direction of the current and the voltage drop appear for  some pairs of contacts. This effect of the ``absolute negative resistance'' was proposed as an evidence of viscous flows of
%\textcolor{red}{the electron fluid}
electrons 
in~\cite{Levitov_et_al} and was observed for graphene samples in~\cite{grahene}.  Second, the dependencies of the electron viscosity on magnetic field and flow frequency
are very specific and can be used to characterize the viscous electron fluid. The giant negative magnetoresistance observed on high-mobility GaAs quantum wells \cite{exps_neg_1, exps_neg_2, exps_neg_3, exps_neg_4} was explained by this effect and  thereby was employed to detect the hydrodynamic transport \cite{je_visc}.  For ac flows of  the  electron fluid, the viscosity exhibits the resonance at the doubled electron cyclotron frequency~\cite{Alekseev_Alekseeva,future}. Such resonance manifests itself in responses of 
its
%\textcolor{blue}{its}
%\textcolor{red}{the electron fluid}
conductance on incident radiation, that was apparently observed in~\cite{exp_ac_GaAs_1,exp_ac_GaAs_2,exp_ac_GaAs_3}. In particular in a strongly non-ideal electron fluid     the ac flow is formed by transverse shear stress waves, whose dispersion law reflects the resonance in the viscosity coefficients.  In recent works~\cite{profile_1,profile_2} direct observations  of the profiles of the Hall electric field and the current density for a Poiseuille flow  of 2D electrons in graphene stripes by means of space resolved measurements of electric and magnetic field  were  reported.

All these methods are quite difficult to use: they require either an analysis of data on a number of specially designed samples with a given geometry or %%%\textcolor{red}{the} 
applying %%%\textcolor{red}{of a}
sufficiently strong magnetic fields. Therefore,  simpler,  weaker-invasive methods are wanted. 
%\textcolor{blue}{highly} wanted. 

%\begin{figure*}[htbp]
	%\centering
	%\setlength{\abovecaptionskip}{-200pt}
%	\includegraphics[scale=0.4]{Scheme_hydro_by_spin.png}%{spin4}%
%\includegraphics[width=0.35\textwidth]{Absorb.pdf}%{spin4}%	
%	\caption{} 
%\label{fig1}
%\end{figure*}

In this work we propose  % come up  with 
a spin-injection-based method  to detect  the hydrodynamic regime of electric transport in an electron fluid (see % the scheme in
Fig.~\ref{fig1}). 
%%%%\textcolor{red}{We focus on materials with weak spin-orbit interaction where the interconnection of the spin and space degrees of freedom is not substantial for the observed transport phenomena.  }
We argue 
%(\textcolor{blue}{argue}) 
that the space distribution of the injected spins in a pure sample can be employed to 
%\textcolor{red}{detect}
%\textcolor{blue}{visualize}
visualize
a viscous flow ("spin imaging technique"). 
% \textcolor{red}{(improve!) 
Namely, % we show that there is a stark contrast 
the distributions of the electron spin density  and  its magnetic field dependence for the  Poiseuille flow of the viscous electron fluid, being inhomogeneous by the section of a sample,  
strongly differs 
from the Ohmic regime.
%%%\textcolor{red}{from the ones for Ohmic flows.}
% The ones  
% \textcolor{blue}{
% For the first  flow with an  almost homogeneous   distribution of the drift 
% velocity across a contact  and homogeneous spin injection, the distribution of % the spin density remains almost homogeneous far from the contact. 
% For the second one with the strongly inhomogeneous parabolic velocity profile, %  $u_x = 4 v_\ast y(W-y)$, 
% at  the drift-dominated regime of the spin current, the spin distribution  
% becomes 
% inhomogeneous far from the contact, following the Poiseuille electric current. % Detecting such  spin distribution together with  the magnetic field 
% dependencies % of spin at different points across the sample section 
% (the desynchronized Hanle % curves) is to be  a direct experimental 
% method attesting to the formation of 
% Poiseuille electronic flow.}
We demonstrate this concept by performing the numerical calculations with realistic parameters for high-mobility samples. 
The advantage of the proposed method is that 
the injected spin distribution 
%%%%%\textcolor{red}{the injected and being homogeneously or inhomogeneously of distributed spin polarization}
provides almost no effect on the magnitude and profile 
% \textcolor{red}{a given hydrodynamic or not hydrodynamic electron flow}
% \textcolor{blue}{(replace by: 
of a given electron flow.
In this way, the proposed technique is "light and non-invasive", 
as compared with the 
already existing methods of detection of the viscous electron fluid. 

%\textcolor{red}{those require samples with complex geometry of contact or
% applied a strong magnetic filed modifying the flow. (grammatically strange)}

We consider a flow of 2D viscous electron fluid in high-mobility  samples, 
%%%%in graphene or GaAs quantum well  injected from a bulk metalic  contact. 
%%%%In high-mobility samples the time of electron scattering on impurities can be  smaller than that corresponding to electron-electron collisions. 
where the time of electron scattering on impurities can be 
longer 
%\textcolor{red}{smaller}
than that corresponding to electron-electron collisions.
In this regime the electron momentum relaxation takes place predominantly or mostly at the channel boundaries with the subsequent formation of the Poiseuille flow. 
%%%In \textcolor{red}{experiments}with electric current flowing through long  and sufficiently wide samples (see, for example, Refs.~\cite{profile_1,profile_2,Gusev_1,Gusev_2,Gusev_3}), 
In case of long  and sufficiently wide samples (realized, for example, in experiments~\cite{profile_1,profile_2,Gusev_1,Gusev_2,Gusev_3}) 
the velocity distribution can be 
found
from the Navier-Stokes  equation.  
The result for low-frequency flows 
%\textcolor{red}{for} 
%\textcolor{blue}{in}
in samples with some density of  disorder and fully rough  edges  takes the form~\cite{Gurzhi}: 
\begin{equation}
\label{eq:ux}
    V_x(y) = \frac{e \, E_x }{m\,\tau_{tr}}  \,\Big\{ 1- \frac{\cosh[(y-W/2)/l_G]}{\cosh[(W/2)/l_G]} \Big\}
    \:,
\end{equation}
where 
%\textcolor{blue}{
$m$ is an electron effective mass, $W$ is the channel width, %}
$\tau_{tr}$ is the momentum relaxation time in the bulk due to scattering of electrons on disorder
or 
%\textcolor{red}{and/or}
phonons,  $l_G = \sqrt{\eta \tau_{tr} } $ is the Gurzhi length,  $\eta = v_F^2 \tau_{ee}/4$ is the  viscosity of the electron fluid determined by the electron-electron collisions time, $\tau_{ee}$ is the time of relaxation of the shear stress due to inter-particle collisions, and  $v_F$ is the Fermi velocity. 
%\textcolor{blue}{
Equation (\ref{eq:ux}) describes 
%%%%turns into 
the homogeneous Ohmic flow $V_{\rm Ohm} =  e \, E_x /( m\,\tau_{tr} ) $ 
at $l_{G} \ll W$,
while the Poiseuille parabolic distribution 
appears in the opposite limit $l_G \gtrsim W $ (see Fig.~\ref{fig1}).

\begin{figure}[t!]
	\centering
	\includegraphics[width=0.45\textwidth]{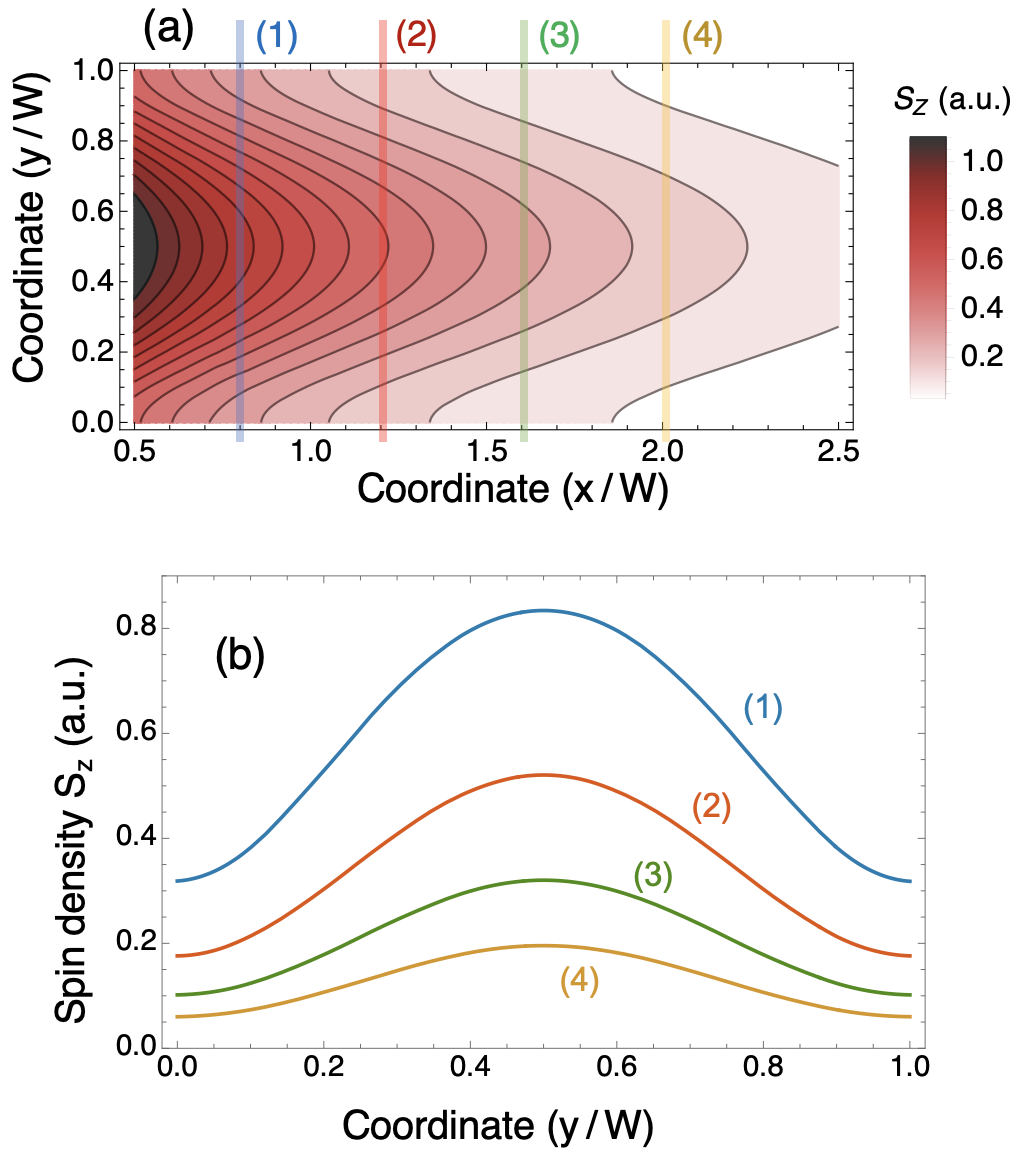}%{spin4}%
	\caption{
	(a): Calculated distribution of 
	viscous electron fluid spin polarization %%%%profile of	viscous electron fluid 
	in the channel. 
	(b): Profiles of the spin density 
	%%%Cross-sections of the profile 
	%%%along $y$-direction 
	at different $x$-positions: 
	$x=0.8W$ (1), $x=1.2W$ (2), $x=1.6W$ (3), and $x=2W$ (4).
The parameters $\tau_s = 200$~ps, $D_s = 60$~cm$^2$/s ($L_s \approx 1.1 ~\mu$m), 
	%%%the drift velocity in the center of channel 
	$V_\ast = 4 \times 10^6$~cm/s ($L_d = 8~\mu$m) and $W=8 \mu$m.
	} 	
	%\textcolor{red}{The calculation results for}
	%the \textcolor{blue}{calculated} spin polarization profile of 
	%\textcolor{red}{electronic flow}
	%\textcolor{blue}{viscous electron fluid}
	%in the channel. All coordinates are in terms of the width of the channel $W$. Panel (a): A 2D contour-map of the %profile. Panel (b): 
	%\textcolor{red}{Some four}
	%cross-sections of the profile along $y$-direction at definite $x$-positions: $x=0.8W$ (blue line), $x=1.2W$ (orange %line), $x=1.6W$ (green line), and $x=2W$ (yellow line).
\label{fig2}
\end{figure}

In this work we address to systems with no pronounced 
%weak spin-orbital effects and neglect its 
%with illegible non-dominant spin As a model here we stick to , 
%Namely, we neglect the possible 
%
%as well as possible boundary spin accumulation due to the spin Hall effect (Glazov). 
effects of spin-orbital 
%interaction 
coupling 
on electric current distribution and neglect the extra-boundary 
spin accumulation due to the spin Hall effect~\cite{Dyakonov__spin_Hall_magnetores, Alekseev_Dyakonov__spin_Hall_magnetoimped}
%\textcolor{red}{in ohmic regime}
or any rotational viscosity effects~\cite{Maekawa_et_al__rotat_visc_1, Maekawa_et_al__rotat_visc_2, Maekawa_et_al__rotat_visc_3, Maekawa_et_al__rotat_visc_4, Polini__rotat_visc}. 
%\textcolor{red}{in hydrodynamic regime of charge transport}
The latter could lead to a vorticity of the electron flow, which induces a torque acting on electron spin and results in generation of the spin density~\cite{Glazov___spin_Hall_in_hydr}. %%%%%\textcolor{red}{thoroughly studied in}~
By other words, our consideration is valid for systems where the electric current profile is settled according to Eq.~(\ref{eq:ux}), while the distribution of the spin density follows the local drift velocity of the electron fluid and does not affect its orbital motion. 
In this approximation the distribution of the spin density $\bm{S}$ can be determined based on the drift-diffusion model~\cite{fabian2007semiconductor}
formulated in the following equation
%Namely, we consider the 
%following equation for the spin evolution
\begin{equation}
\label{eq:spin}
    %\frac{\p \bm S}{\p t} 
    \dot{\bm S}
    + \nabla_i\bm q^i = \left[ \bm{\omega}_c \times \bm{S} \right]  - 
    \bm{ S}/ \tau_s
    %\frac{\bm S}{\tau_s}
\end{equation}
where $\bm q^i$ is the spin current (the flow of the value $\bm S$ along  the direction $x_i$), $\bm{\omega}_c$ is the Larmor precession frequency due to the in-plane magnetic field, and $\tau_s$ is the spin relaxation time, which is assumed to be isotropic. 
%Within the drift-diffusion model it contains two contributions
The spin current $\bm q^i$ contains two contributions
\begin{equation}
\label{eq:spinCurr}
    \bm q^i = - D_s \nabla_i \bm S + V_i \bm S,
\end{equation}
where the first term describes the spin-diffusion with coefficient $D_s$
and the second term stems from the 
%motion 
%\textcolor{blue}{drag}
drag 
of the spin density with the drift velocity $\bm{V} = \bm{e}_x V_x(y)$ determined by Eq.~(\ref{eq:ux}). 
%in the electric field. 
%%%%is the drift with the velocity  of the spin density due to stems from the electric field $\bm E$ induced drift velocity $\bm u$. 
%Naturally, 
%In case of the Poiseuille flow the drift velocity is inhomogeneous across the channel coordinate. 
For further consideration we also use  
%%%In what follows it is convenient to use two spin-related parameters, namely 
the spin diffusion length $L_s = \sqrt{D_s \tau_s}$ and the spin drift length $L_d = V_\ast \tau_s$ in the centre of channel,
%\textcolor{blue}{
$V_\ast = V_x(W/2)$.

%Some discussion of the modifications in $D_s, \tau_s$ due to electron-electron collisions.
%\textcolor{blue}{To Kirill: maybe to improve this part and to add Refs?}
%Let us discuss the specifics of the spin transport in high-mobility 2DEG upon frequent electron-electron collisions. 
%%%point out that fast electron-electron collisions specific to the considered hydrodynamical regime affect significantly the magnitude of $D_s, \tau_s$.  
%Let us discuss several important issues related to the spin transport and 
%specific to the considered hydrodynamical regime. 
%It is important to note that the fast electron-electron collisions affect significantly the magnitude of $D_s, \tau_s$.
In general, the spin diffusion length in high mobility samples can be extremely large due to long momentum relaxation times $\tau_{tr}$.
%%%on impurities $\tau_i$. 
For instance, in clean graphene-based lateral spin valves the spin can diffuse to a distance of up to $1~\mu$m~\cite{drogeler2014nanosecond,guimaraes2012spin,ahn20202d}. 
%diffusion length $L_s$ can be as high as.
However, when the viscous electron fluid is formed, 
%%%for the viscous electron fluid considered in this paper 
%\textcolor{blue}{
$D_s$ is significantly reduced~\cite{angel2018field} as it is mostly govern by electron-electron scattering,
%%%\textcolor{red}{the length $L_s$ of diffusive spin transport is mostly govern by electron-electron scattering, thus it significantly reduces~\cite{angel2018field},} 
herewith 
%%the magnitude of  can be significantly reduced , 
%%%while 
%as the $L_s$ 
% \textcolor{blue}{the spin relaxation time $\tau_s$ controlled by % D'yakonov-Perel-like mechanism is also modified 
% $1/ \tau_s \sim  \Omega_{so}^2 \tau_{ee}$ % \cite{Glazov_Ivchenko_2,leyland2007enhanced}.
% }
% \textcolor{red}{
the spin relaxation time $\tau_s$ is controlled by 
%%%%gets mostly govern by 
%%%%electron-electron scattering via 
D'yakonov-Perel-like mechanism giving $1/\tau_s \sim  \Omega_{so}^2 \tau_{ee}$ \cite{Glazov_Ivchenko_2,leyland2007enhanced}.
% }
%rather than the impurities. 
%%%$\tau_{ee}$ instead of $\tau_i$. 
The decrease of $L_s$ due to the inter-particle scattering  is %appears to be 
favorable for the spin imaging of the Poiseuille flow, 
as the weakening of the spin diffusion prevents distortion of an inhomogeneous spin pattern.
%this will be explained in more detail hereinafter. 
%In particular, the spin diffusion coefficient becomes significantly reduced compared with the estimation based on impurity-scattering. 
%%In ultra-high mobility samples the latter can be extremely large, so can be the spin diffusion coefficient, for instance in clean graphene-based lateral spin valves the spin diffusion length $L_s$ can be as high as.
%%The systems in our consideration, though keeps , is strongly affected by the electron-electron collisions, thus the gets much smaller, see e.g.~\cite{angel2018field}. 
%Also, the spin relaxation can be sensitive to electron-electron scattering. 
%\textcolor{blue}{some short discussion}
%An example is the D'yakonov-Perel mechanism 
%Note that in high-quality samples with long $\tau_i$ the very approximation of the spin relaxation fails, as the spin density exhibits more complex oscillating dynamics (Refs). 
%. Without effective one should expect the oscillating decrease, while effectively replaces $\tau_s^{-1} = \Omega_{so}^2 \tau_{ee}$.
%The , the spin relaxation terms are sensitive, see the details in Ref(Glazov-200?, Glazov,Ivchenko-200?). 

\begin{figure}[t!]
	\centering
	\includegraphics[width=0.45\textwidth]{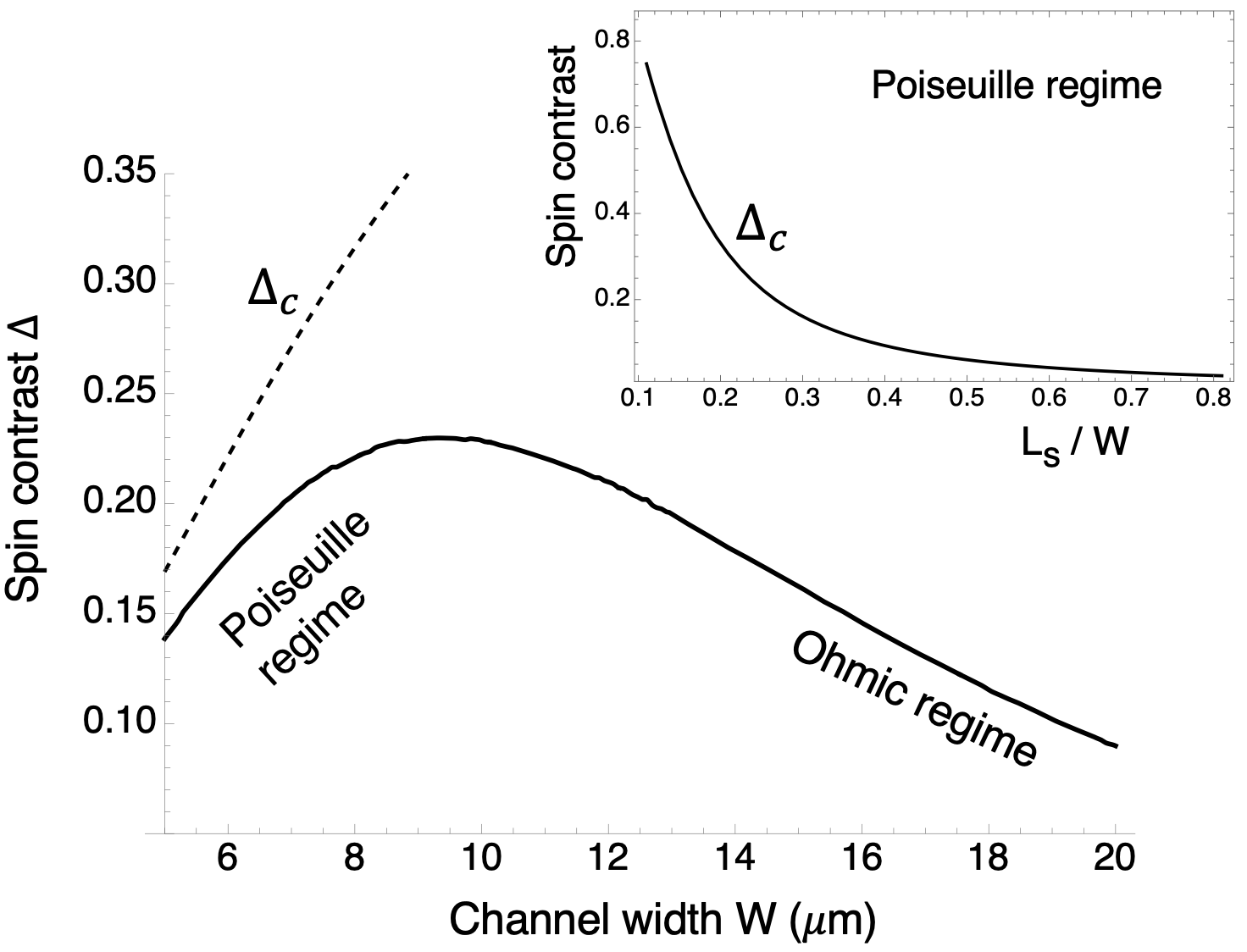}%{spin4}%
	\caption{The dependence of spin polarization contrast across the channel on its width $W$.
	The approximation $\Delta_c$ from Eq.~(\ref{eq:tail}) is shown by the dotted curve. 
	%the change of channel width $W$. 
	%%%%%%At small $W$ the Poiseuille regime matters and there is a growth of the contrast due to the decrease of the ratio $L_s/W$, which makes electrons to prevent their spin polarisation (this tendency is shown on the Panel). 
	The inset shows the dependence $\Delta_c$ on $L_s/W$ for the Poiseuille regime. % and is 
	%%%%%At higher $W$ the transition to the ohmic regime occurs, and hence there is a decrease of the spin contrast due to the planarisation of the velocities profile.
	} 
\label{fig3}
\end{figure}

%%%We proceed with discussing the spin injection in the electronic hydrodynamical channel at zero magnetic field. 

%In  we demonstrate 
The distribution of spin density $S_z$ 
emerging due to the Poiseuille flow of electrons in the hydrodynamical regime at $\omega_c=0$
is demonstrated in Fig.~\ref{fig2}(a). 
%%%used in Fig.~\ref{fig2} 
%%%%%The parameters are taken for high-mobility GaAs quantum wells~\cite{angel2018field}, the channel width $W=8 \mu$m.  
The parameters are relevant for high-mobility GaAs quantum wells~\cite{angel2018field,leyland2007enhanced} and are described in the caption to Fig~\ref{fig2}, the channel width $W=8\; \mu$m.  
%%%%%%%%%%namely, we consider the channel width $W=8 \mu$m typical for hydrodynamical experiments (Refs) and take $\tau_s = 200$~ps, $D_s = 60$~cm$^2$/s ($L_s \approx 1.1 ~\mu$m), the drift velocity in the center of channel $v_\ast = 4 \times 10^6$~cm/s corresponding to $L_d = 8~\mu$m. 

As boundary conditions we used the absence of spin current $q_y^z=0$ at $y=0,W$, 
%%%As for the boundary conditions we implemented the absence of spin current $q_y^z=0$ at $y=0,W$, 
we also assumed that $q_x^z$ is proportional to the electric current flowing through the contact at $x=0$~\cite{fabian2007semiconductor}. 
%%%We also implemented the boundary condition at $x=0$ used in spin injection schemes (Refs), i.e. 
%%%%%%the spin injection condition at $x=0$~\cite{fabian2007semiconductor}; the latter suggests that the spin current is proportional to the electric current flowing through the contact. 
In our case we take ${q}_x^z \propto V_x(y)$ being consistent with the %Poiseuille 
hydrodynamic 
distribution of the electric current in the bulk of the channel. 
%across the channel. 
We recognize that the electric current distribution right in the vicinity of a contact can 
alter from the Poiseuille form. %%% to some extent. 
%%%%%%particular spin distribution nearby the injector can be slightly altered from the calculated pattern due to some unaccounted perturbations in the boundary conditions and drift velocity spreading. 
However, beyond some transition region (no longer than $x \lesssim W$)
%\textcolor{blue}{
the drift velocity will be settled according to Eq.~(\ref{eq:ux}) and
%the spin distribution 
$S_z$ will keep strongly inhomogeneous shape
due to an effective dragging by $\bm{E}$ in the center of the channel. 
%%%determined 
%%%the transverse diffusion spreading of the dragged spin density in the center of the channel
%%by the transverse diffusion of the spin density dragged by $\bm{E}$ in the center of the channel .
%}

%\textcolor{red}{the balance between transverse spin diffusion and longitudinal drift of the spin density in the center of the channel. }
%%%longitudinal drift component of the spin current. %%%%% in the core.

%%%%%%%Note that the behavior of $S_z$ away from the injector cannot be significantly altered due to some unaccounted perturbations in the boundary conditions as it is mainly determined by the interplay between Poiseuille distribution of drift velocities and spin diffusion in the bulk.
%%%%%nearby the contact region,  
%%while some unaccounted perturbations nearby the contact region 
%can emerge , 
%the behavior of the spin density outside the transition region cannot be significantly altered due to 
%these perturbations 
%some perturbations nearby the contact region
%it is mostly determined by the interplay between Poiseuille distribution of drift velocities and spin diffusion in the bulk.
%so the considered scheme is for analyzing boundary condition seems justified
%is minimal model. 
%does not depend on the details nearby the contact region 
%and remains mostly determined by the interplay between Poiseuille distribution of drift velocities and spin diffusion. 

This feature is clearly seen in Fig.~\ref{fig2}(b), 
where we demonstrate the profiles of the spin density across the channel 
at different distances from the left boundary. 
The tail of the spin distribution presented in Fig.~\ref{fig2}(b) 
can be approximated by the expression
\begin{equation}
    \label{eq:tail}
    S_z(x,y) = A\, e^{-x/L_{\ast}} \left[ 1 - \Delta_c \cos{
    \left(
    \frac{2\pi y}{W}\right)
    } \right],
    %= \mathcal{S}(x) \left( 1 - \delta \cos{\left(2\pi y/W\right)} \right),
\end{equation}
where $A$ is a constant being independent of $x$ and $y$, 
the factor 
$e^{-x/L_{\ast}}$ 
%$\mathcal{S}(x) \propto e^{-x/L_{\ast}}$ 
describes the spin density decay 
with an averaged drift length $L_{\ast} = \langle V_x \tau_s \rangle = 2 L_d/3 $ and 
%%%%at characteristic distance $L_{\ast} = \langle u_d \tau_s \rangle = 2 v_{\ast} \tau_s/3 $ determined by the average drift length. 
%is an averaged electron spin across the channel 
%%%and the parameter $\delta$ depends on $L_s/W$. 
%depends on $\tau_s, D_s, W$. 
the parameter $\Delta_c$ determines the contrast of the spin imaging. %%%%%, it is thus of key importance.  
%%%%%Indeed, to visualize an inhomogeneous drift velocity profile across the channel one needs to have a noticeable variation of the spin density, the latter being determined by $\delta$. 
%Thus, the contrast parameter $\delta$ is of key importance and its . 
%%with the increasing effectiveness of spin diffusion along $y$-direction the spin contrast $\delta \to 0$ will vanish. 
%magnitude of $L_s \sim W$ the contrast will diminish due to the effective spin diffusion along $y$-direction. 
%it is  at larger $\delta$. 
%the larger $\delta$ the . 
%In fact, it is the parameter $\delta$ that controls the contrast of the spin imaging. 
%In particular, to visualize the Poiseuille flow 
%the inhomogeneous spin texture must not be destroyed by the spin diffusion along $y$-direction; 
%by other words $\delta \to 0$ and the imaging technique loses sense at $W \lesssim L_s$. 
%\textcolor{blue}{
%Visualization 
It follows from Eq.~\ref{eq:tail} that 
imaging
of the Poiseuille flow by spin injection is possible provided that 
%%%there is a noticeable magnitude of 
%spin contrast.
$\Delta_c \sim 1$.

%%%%%%%Note that the spin contrast $\delta \to 0$ must vanish if the spin diffusion along $y$-direction is rather effective. 
%%%if the spin diffusion along $y$-direction is quite effective, the spin contrast $\delta \to 0$ will vanish. 
%%%%%To determine the critical region beyond which there is no spin imaging of the Poiseuille flow 
%To analyze more carefully the decrease of $\delta$ on spin parameters 
%the expression for $\delta$ we using biharmonic approximation. 
%%we get an approximated analytical solution for Eq.~\ref{eq:spin}. 
%For the spin imaging of Poiseuille flow %drift velocity profile 
%\textcolor{red}{To visualize the Poiseuille flow, meaning to get noticeable magnitude of spin contrast $\delta$, an inhomogeneous spin density originating from $y$-dependent drift velocity must survive spreading due to spin diffusion in $y$-direction.  }
%For the spin imaging it is crucial that an inhomogeneous spin density originating from $y$-dependent drift velocity survives averaging due the spin diffusion along $y$-direction.  

To determine the critical region beyond which $\Delta_c$ vanishes 
%disappears 
%%the contrast 
%$\delta \to 0$ 
%the spin contrast $\delta \to 0$ vanishes  
we get an approximated analytical solution for Eq.~(\ref{eq:spin}). 
We use the biharmonic approximation for $y$-dependence of $S_z(x,y)$ and $V_x(y)$ and 
take into account only the drift component of $q_x^z$. 
%consider only the slowest decaying solution at the tail of the spin density; 
%\textcolor{red}{for which }
For the slowest decaying solution at the tail of the spin density
we obtain $\Delta_c = 6/[ \, 2\pi^4 \xi^2 + \sqrt{ 18 + 72 \pi^2 \xi^2 + 4 \pi^8 \xi^4}\,]$, 
%the following expression for $\delta$
%We analyzed the mathematical structure of solutions and found out the exact expression for $\delta$ to be 
%\begin{equation}
%\label{eq:contrast}
%    \delta = \frac{6}{ 2\pi^4 \beta^2 + \sqrt{ 18 + 72 \pi^2 \beta^2 + 4 \pi^8 \beta^4}}, 
%\end{equation}
where $\xi = L_s/W$. 
The expression Eq.~(\ref{eq:tail}) with $\Delta_c$ from above is justified in range $1 \lesssim  \xi \lesssim 0.05$, where the right boundary is determined by the failure of biharmonic approximation due to a 
weakened spin diffusion. 
%very  weak spin diffusion. 
%%%%%in this condition corresponds to $\delta \to 1$ when biharmonic approximation fails (the spin diffusion is weak).
%%%%%parameter $\beta$, 
The spin distribution plotted in Fig.~\ref{fig2} have the parameters $\xi = 0.14$, $\Delta_c = 0.52$ and the expression from Eq.~(\ref{eq:tail}) fits well with the numerical solution starting from $x /W\gtrsim 1.2$.

%%%%\textcolor{blue}{New paragraph with the spin contrast including Gurzhi}
The dependence of  $\Delta_c$ on the ratio $L_s/W$ is shown on the inset of Fig.~\ref{fig3}. 
%\textcolor{red}{It follows from the obtained expression that}
It is seen that 
 %for $\delta$ that  
%One feature of 
the spin contrast decreases significantly 
$\Delta_c \lesssim 0.2$ already at $\xi \gtrsim 0.3$, instead of $\xi \gtrsim 1$ as might be expected.   
%%%%%%%%According to the found estimation the spin contrast decreases significantly $\delta \lesssim 0.2$ already at $\beta \gtrsim 0.3$, instead of $\beta \gtrsim 1$ as might be expected.   
Since $\Delta_c$ is determined by single ratio $L_s/W$, 
the applicability of the spin imaging approach 
for sufficiently large in-plane electric fields (when  $L_d \gg L_s$) 
is govern by a simple criteria $W \gtrsim 3 L_s $,
%%%%%%%Interestingly, as $\delta$ is determined by single ratio $L_s/W$ which suggests that for sufficiently large in-plane electric fields (realizing $L_d \gg L_s$) the applicability of spin imaging approach is govern by a simple criteria $W > 3 L_s $. 
%%%$L_s/W \lesssim 0.3$. 
%Naturally, 
suggesting that a more plausible situation is realized for sufficiently wide samples. 
%%%%However, there is a general limitation for the spin imaging technique at large channel widths, which is connected with the breaking of the condition $W \ll l_G$. 
%%%Indeed, according to Eq.(\ref{eq:ux}) for the electric current density for rather wide samples the Poiseuille flow transforms into the homogeneous Ohmic regime (with no spin imaging structure). 
The spin contrast across the 
transition between Poiseuille and Ohmic transport regimes 
is illustrated in Fig.~\ref{fig3}, 
where we present the dependence of 
$\Delta = [\, S_z(y_c)-S_z(y_e) \, ]/[ \, S_z(y_c)+S_z(y_e)\, ] $ on $W$; 
here 
%%%%the spin contrast $\delta'$
%%determined as $ (S_z(W/2)-S_z(0.1 W))/(S_z(W/2)+S_z(0.1 W)) $ 
%%%on the channel width with account for the transition of drift velocity from Eq.~(\ref{eq:ux}). The parameter $\delta' = [\, S_z(y_c)-S_z(y_e) \, ]/[ \, S_z(y_c)+S_z(y_e)\, ] $ 
$y_c=W/2$ is in the 
center of the channel and 
 $y_e= W/10$ is nearby its boundary. 
A nonmonotonic character of $\Delta$ stems from the fast
initial increase of $\Delta_c(L_s/W)$ relevant for the Poiseuille flow,
which is further suppressed at larger $W$ due to the transition to the Ohmic regime.

%\onecolumngrid
%\begin{center}
\begin{figure*}[t!]
	\centering
\includegraphics[width=0.9\textwidth]{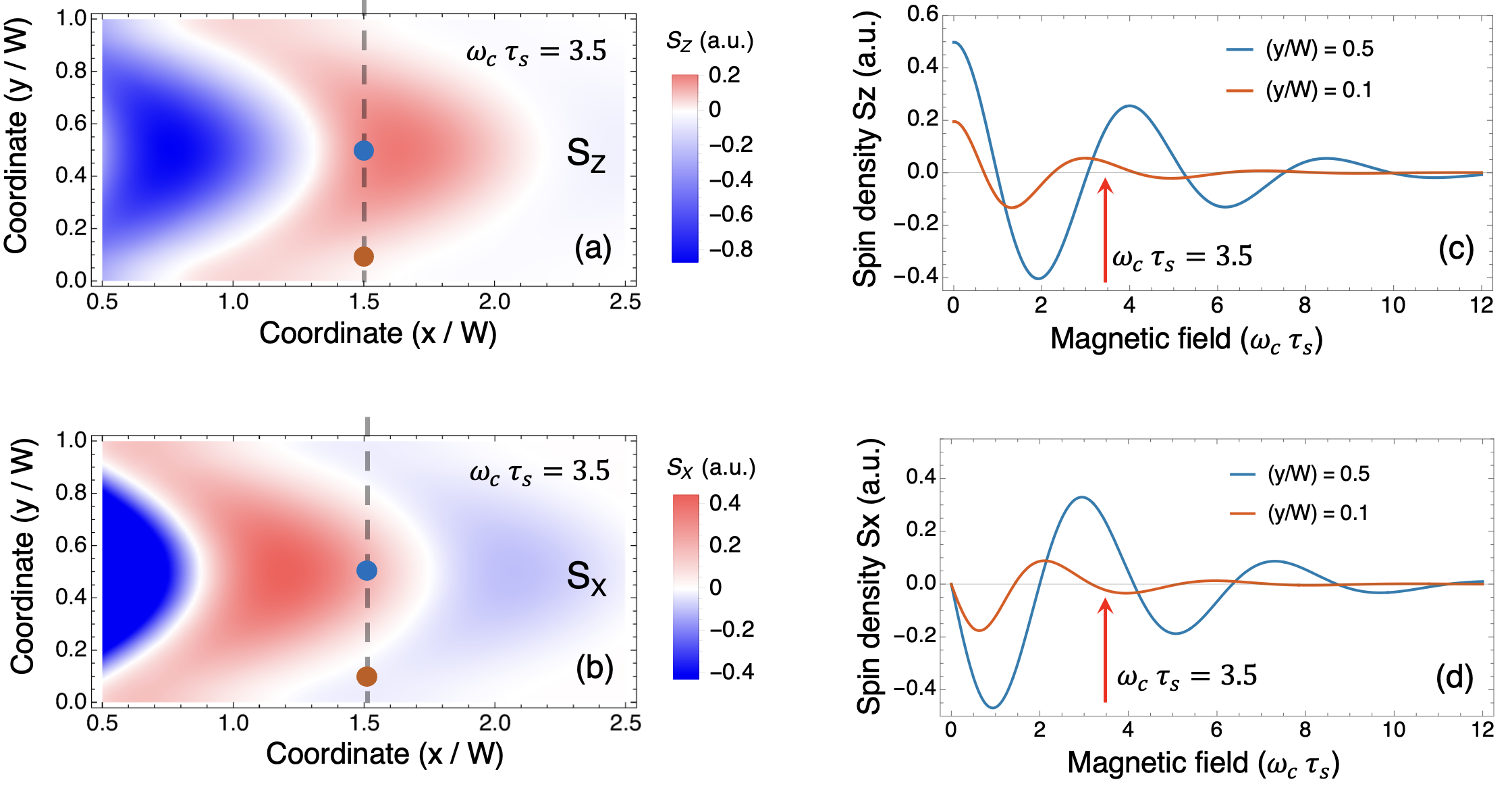}%{spin4}%
	\caption{
	%Local spin components of electrons in the channel. 
	Distribution of the spin density in magnetic field directed along the $y$ axis. 
	(a,b): Space-resolved densities $S_z$ and~$S_x$  at $\omega_c \tau_s = 3.5$. %, respectively. 
	(c,d): 
	Magnetic field dependences of $S_z$ and $S_x$ calculated at $x = 12 \;\mu$m away from the injector.  
	%%%when the constant magnetic field is switched on along the $y$ axis. 
	The Hanle curves on the right panels are taken at the center ($y=0.5W$, blue color) and near the edge ($y=0.1W$, orange color) of the channel. The positions of spin detection in each case are depicted by big dots of 
	corresponded colors on the left panels. % at $\omega_c \tau_s = 3.5$. 
	%%%which are calculated for the case when the condition $\omega_c \tau_s = 3.5$ holds.
	} 
\label{fig4}
\end{figure*}

Another spin-related evidence  % smoking gun
of the Poiseuille flow of the electron fluid 
revealed by the spin imaging technique is 
the appearance of nonzero relative phase shifts for the Hanle curves (the dependence of $S_{z,x}$ on $\omega_c \tau_s$) probed at different positions across the channel.
%%%%%%the desynchronization of the Hanle curves (the dependence of $S_{z,x}$ on $\omega_c \tau_s$) across the channel. 
%Add physical explanation based on basic formulas for the Hanle curves upon spin injection. 
Below we explain  this idea in detail. 
We keep to the drift-dominated regime, 
at that the difference between two Larmor frequencies $\Delta \omega_c$ at which 
the Hanle curve exhibits the neighboring peaks or dips at a fixed point of spin probe in space 
%at $x_0$ distance of the spin probe 
%the period of the Larmor frequencies for which the spin repeats and peaks and dips 
%spin probe point in space $x_0$  
%at which the spin repeats and peaks and dips are expected 
can be estimated as $\Delta \omega_c = (2\pi V_x)/x_0$, where $x_0$ is the distance from the injector. 
%Naturally, 
In case of the Poiseuille electronic flow the drift velocity $V_x(y)$ from Eq.~(\ref{eq:ux}) changes significantly across the channel leading to the variation of $\Delta \omega_c$. 
%%%%%distribution of drift velocity the typical period is different across the sample leading to the variation of $\Delta \omega_c$. 
An ultimate manifestation of this feature would be the shift $\Delta \omega_c$ as a function of $y$-coordinate at fixed distance $x_0$
resulting in the ``desynchronization'' of the Hanle curves.

We proceed with considering this scenario in more detail. 
We apply an in-plane 
%%%%We chose the direction of the 
magnetic field $\bm{\omega}_c = \omega_c \bm{e}_y$ and keep only the injection of $S_z$
%%%%$z$-component of the spin density 
(the boundary conditions are the same as Fig.~\ref{fig2}; the spin current $q_x^x=0$ is absent). 
In Fig.~\ref{fig4}(a,b) we demonstrate 
inhomogeneous oscillating spatial patterns of $S_{z}, S_x$
%the spin densities $S_{z}$, $S_x$
%%the colored map of the spin densities $S_{z}$, $S_x$ distribution 
%distribution of $S_{z}$ and $S_x$ 
inside the electronic channel at fixed magnetic field $\omega_c \tau_s = 3.5$ (red and blue colors stand for the positive and negative signs, respectively), other parameters are the same as in Fig.~\ref{fig2}. 
%\textcolor{blue}{Do we need to specify that red and blue colors stand for the positive and negative signs, respectively?}
%giving an rigorous description of the spin distribution emerging in finite magnetic field. 
%We chose the direction of the magnetic field $\bm{\omega}_c = \omega_c \bm{e}_y$ and 
%To investigate this feature 
%numerical simulation of Eq.~\ref{eq:spin} 
%In Fig.~\ref{fig3}(a,b) we present the colored distribution of $S_{z,x}$ inside the electron channel a fixed magnitude of the magnetic field $\omega \tau_s = 3.5$ (the direction ) 
%We demonstrate these ideas in Fig.~\ref{fig3}, where we present the spin distribution in the channel for $\omega \tau_s = 3.5$; 
%the parameters are the same as in Fig.~\ref{fig2}. 
%%%It follows from Fig.~\ref{fig3} that $S_{z,x}$ possess an inhomogeneous oscillating spatial pattern.  
%%due to the precession in the transversal magnetic field. 
The procession structure of the spin density visible in Fig.~\ref{fig4} is specific for the drift-dominated regime of the spin transport~\cite{fabian2007semiconductor}. 
%%%%and serves in experiments to identify it. 
Importantly, $S_{z,x}$ are sign-altering and 
%%%sign-altering density $S_{z,x}$ remain 
inhomogeneous at the same moment, %there is a pronounced core at the center of the channel; 
one can clearly observe that at fixed distance $x_0$ %from the injection 
the spin density can have different signs in the center and nearby the boundary of the channel. 
This feature is explicitly connected with the discussed variation in drift velocities 
emerging for the hydrodynamic regime of viscous electron fluid transport. 
%%%%%of the electron fluid Poiseuille flow. 
%%%hydrodynamic flow
%, indicating the 
%in addition to an inhomogeneous structure with a pronounced core at the center of the channel,
%the spin densities $S_{z,x}$ also exhibit an oscillating structure due to the precession in the transversal magnetic field. 
%%%%%that the spin distribution preserves an inhomogeneous structure with a pronounced core at the center of the channel and, in addition, while exhibiting an oscillating structure due to the precession in the transversal magnetic field. 
%%%This is exactly the revelation of the . 

%\textcolor{red}{(ELIMINATE) In Fig.~\ref{fig4}(c,d) we demonstrate the Hanle curves %(the dependence of $S_{z,x}$ on $\omega_c \tau_s$) 
%%%is explicitly seen in the Hanle curves plotted in Fig.~\ref{fig3}(c,d). 
%%%This is explicitly demonstrated in Fig.~\ref{fig3}(c,d) where we present the Hanle curves 
%calculated at two spatial positions with transverse coordinate $y/W = 0.5,0.1$ 
%(blue and orange dots in Fig.~\ref{fig4}(a,b), respectively) 
%located $x = 12 \;\mu$m away from the injector.} 
%.(blue and orange dots in Fig.~\ref{fig3} corresponding to $y=(0.5,0.1) W$). 
%%contacts across the channel.  
The relative shifts of the oscillations period $\Delta \omega_c$ of the Hanle curves are shown in Fig.~\ref{fig4}(c,d).  The Hanle curves are calculated at two spatial positions across the channel, see Fig.~\ref{fig4}(a,b).
%%%is clearly visible and 
When $\omega_c \tau_s \sim 4$, these shifts are up to $\pi$. 
We argue that the presented desynchronization of the Hanle curves upon hydrodynamical response 
can be used to confirm independently the formation of the viscous electron fluid. 
%due to the hydrodynamical response. 
%%%%%to serve to confirm the formation of viscous electron liquid featured by hydrodynamical response. 
%%%%%%, the shift of the oscillations period at $\omega_c \tau_s \sim 4$ is up to $\pi$, meaning that is positive, while next to the channel boundary it can be negative at the same moment. 
%%%As explained above, we attest to the desynchronization of the oscillation periods for different contacts across the channel. 
%%%%%which can be used to independently evidence the formation of the viscous electron fluid.

%%%as an independent evidence for the formation of the viscous electron fluid.
%which serves as an independent evidence for 
%in the Poiseuille regime, 
%which serves as an independent evidence for 
%the formation of inhomogeneous hydrodynamic flows of electron fluid. 

%Discussion of material systems and the applicability of the spin imaging method.

%%%\textcolor{blue}{Summarizing, the outlook}
%%\textit{Conclusion.}
In conclusion, we have proposed an approach to visualize hydrodynamic electronic flows 
by measuring the spin polarization distribution across the transport channel. 
%%%%imaging by the measuring of the spin polarization distribution across the high-mobility transport channel. 
%%%%The idea is based on interplay of spin polarization and electronic velocities distribution across the channel in Poiseuille-like flows.
Our calculations show that measuring both the spin polarization contrast and Hanle curves 
at the center and at the boundary of the channel 
allow one to disclose the hydrodynamic regime. 
%%%%can easily 
%%%reveal the hydrodynamic regime of electronic flow. 
We believe that the proposed method paves the way towards 
non-invasive studies of hydrodynamic viscous electron fluids in  
samples of different geometry and microscopic structure. 

\textit{Acknowledgements.}
This work has been supported by the Russian Science Foundation (Project  18-72-10111).
P.S.A. thanks the Theoretical Physics and Mathematics Advancement Foundation "BASIS".

\bibliography{Ref}

\end{document}